\documentclass[aps,floatfix,twocolumn,footinbib,amsmath,amssymb]{revtex4}
\usepackage{amsmath,amsthm,amssymb,color,psfrag}
\input epsf.tex
\usepackage{latexsym}
\usepackage{graphicx}

\def\rd{\text{d}}

\def\in{\text{in}}
\def\qcd{\text{QCD}}
\def\scet{\text{SCET}}
\def\out{\text{out}}
\def\gcusp{\Gamma_\text{cusp}}

\def\mus{\mu_s}
\def\muj{\mu_j}
\def\muh{\mu_h}
\def\muw{\mu_\omega}
\def\Rf{\frac{R}{1-R}}

\def\tom{\tau_\omega}
\def\tAo{\tau_{A_1}}
\def\tAt{\tau_{A_2}}
\def\tAq{\tau_{A_q}}
\def\cdt{ \hspace{-0.1em} \cdot \hspace{-0.1em} }

\newcommand{\qbar}{{\bar{q}}}
\newcommand{\e}{\epsilon}
\newcommand{\nbar}{\bar{n}}

\newcommand{\nn}{\nonumber}

\begin{document}

\title{Resummation of jet mass with and without a jet veto}
\author{Randall Kelley}
\author{Matthew D. Schwartz}
\affiliation{Department of Physics, Harvard University, Cambridge, Massachusetts 02138, USA}
\author{Hua Xing Zhu}
\affiliation{Department of Physics and State Key Laboratory of Nuclear Physics and Technology, Peking University, Beijing 100871, China}

\begin{abstract}
Calculating the distribution of jet masses in high-energy collisions
is challenging because fixed-order perturbation theory breaks down near the peak region,
and because multiple scales complicate the resummation.
To avoid using a jet veto, one can consider inclusive observables, in which every particle is in a jet.
We demonstrate that calculating the mass of the hardest jet in multijet events 
can be problematic, and we 
give an example of an inclusive observable, asymmetric thrust, which can be resummed to next-to-next-to-leading
logarithmic accuracy. 
Exclusive observables with out-of-jet regions are more complicated.
Even for $e^+e^-$ dijet events at energy $Q$, to calculate the mass $m$ of jets of size $R$,
one must impose a veto on the energy $\omega$ of extra jets to force dijet kinematics;
then there are both $\log m/Q$ and $\log m/\omega$ singularities.
To proceed, we suggest a refactorization of the soft function in the small $R$ limit
To justify this refactorization, we show that the expansion of the resummed distribution is in excellent agreement
with fixed order. This motivates considering the expansion around small $R$ as a useful handle on
producing phenomenologically useful resummed jet mass distributions. The strong evidence we give for refactorization at
small $R$ is independent of non-global logarithms, which are not the subject of this paper.
\end{abstract}

\maketitle

\section{Introduction}
The Large Hadron Collider has already produced enormous numbers of high energy jets. These jets provide a wealth of information
about QCD as well as a vital area for exploration of new physics.
While much activity has been devoted to the calculation
of the distribution of jets, for example using leading order (LO) or next-to-leading order (NLO) perturbative calculations in QCD,
very little has been said with any precision about the substructure of the jets themselves.
This is unfortunate, because the substructure of jets may be critical to finding new physics.
In fact, there has been much progress over the last few years
on practical methods using jet substructure to separate signal from background at the LHC~\cite{Butterworth:2008iy,Kaplan:2008ie,Abdesselam:2010pt}.
However, even for the simplest of jet shapes, the jet mass, these studies are forced to rely on the Monte Carlo approximation.
Needless to say, it would be great to have a systematically improvable way to calculate jet shapes, and it may even prove
essential if new physics shows up in one of these channels.

The main theoretical difficulty with calculating
substructure is that fixed-order perturbation theory is in general a very bad approximation. For example, at LO,
jet mass distributions for small mass diverge as $\rd \sigma \sim 1/m^2$, while the measured distributions have a peak
 at small mass and then  go to zero. The physical peak structure is easily explained qualitatively as being due to
large logarithmic terms $\alpha_s^i \log^j(m^2/E^2)$, which dominate over the fixed-order expansion in $\alpha_s$
for small enough mass.
The leading large logarithms (LL) are resummed with the Monte Carlos (or analytically~\cite{Catani:1992ua});
however, until recently, it was impossible to go beyond this order. The difficulty stems from the fact that
while the leading Sudakov double logarithms are universal, subleading logarithms are not.
The coefficients of the subleading logs depend on many variables
besides the mass, including the jet size, the jet algorithm, and the directions and energies of the other jets.
Sorting out all these effects and all the scales is a daunting task.
However, with recent advances in effective field theory, it is now possible
to make systematically improvable calculations of jet substructure,
at least in some kinematical regimes.

While we are mainly interested in jet masses at hadron colliders,
from a theoretical perspective it is instructive to start
with $e^+e^-$ machines.
In $e^+e^-$ collisions, jet shapes can be approached by first studying hemisphere jets and inclusive event shapes
like thrust.
The factorization theorem in Soft-Collinear Effective Theory (SCET) writes the doubly differential hemisphere mass
distribution as a convolution of a hard function, describing the short distance $e^+e^- \to q\bar{q}$ process,
jet functions, describing the collinear degrees of freedom in the hemisphere jets, and a soft function, which describes radiation
which can connect the jets to each other~\cite{Fleming:2007qr,Schwartz:2007ib}:
\begin{multline}
 \frac{1}{\sigma_0} \frac{\rd^2 \sigma}{\rd m_L^2 \rd m_R^2} =H(Q^2,\mu) \int \rd k_L \rd k_R \\
\times  J(m_L^2 - k_L Q) J(m_R^2 - k_R Q) S_{\text{hemi}}(k_L, k_R, \mu) \label{mmfact} \,,
\end{multline}
Here $m_L$ ($m_R$) is the invariant mass of the 4-vector sum of the momenta of particles in the left (right) hemisphere,
with respect to the thrust axis, and $\sigma_0$ is the born cross section for $e^+e^-$ to dijets. The challenge which we address in this paper is to modify this result to
describe jets of arbitrary size $R$ while still resumming all of the large logarithms.

Before considering non-hemisphere jets, a few comments are in order about non-global logarithms (NGL).
Non-global logarithms~\cite{Dasgupta:2001sh} arise in situations where there are multiple scales.
They become relevant, for example, when one integrates the double differential mass distribution
over $m_R^2$ to produce the left-hemisphere mass distribution.
The full QCD calculation
of $\rd \sigma/\rd m_L^2$ involves integrals
over hard emissions into the right side which are not in the soft or collinear sectors present in the leading
SCET Lagrangian. To account for these non-global logs in SCET would require incorporating 3-jet and higher
order operators in SCET (see~\cite{Bauer:2006mk,Bauer:2006qp}). These logs appear very difficult to resum
since there is no known simplification for multiple hard emissions in QCD. However, they can be avoided
by carefully chosen observables. For example, thrust and heavy jet mass do not have the non-global
problems that left-hemisphere or light-jet mass have.

There are other types of non-global logs that are relevant even for an observable which is well described
by the leading operator in SCET.
For example, consider the doubly differential jet mass distribution in the regime $m_L \ll m_R \ll Q$.
In this regime, the above factorization formula is
valid because hemisphere masses ($m_L$ and $m_R$) are small. This is in contrast to the left-jet mass,
where in integrating the right-jet mass to its kinematic limit one enters the regime with $m_R \sim Q$ where
the factorization formula fails.
So when $m_L, m_R \ll Q$, the exact distribution is completely dominated by the IR degrees of freedom included in SCET.
However, SCET {\it does not} guarantee that the logarithms of $m_L$ and $m_R$ can be
resummed using renormalization group evolution (RGE).
In particular,
logs of the form $\log(m_L^2/m_R^2)$ are not completely fixed by the RGE.
If by brute force one could calculate to all orders in SCET,
the result would reproduce the $m_L$ and $m_R$ dependence of full QCD when both masses are
small~\footnote[1]{After the first version of this paper appeared,
the two-loop hemisphere soft function was calculated exactly~\cite{Hornig:2011iu,Kelley:2011ng},
showing that SCET does indeed reproduce the singular behavior of QCD when $m_L \ll m_R \ll Q$.}.
In this paper, we do not attempt to predict or resum these challenging NGLs. Instead, we give evidence that a certain soft function refactorizes in the small $R$ limit. This
provides an orthogonal direction in the study of jet shapes to the investigation of NGLs.
%Resumming this type of non-global logs may be achievable in SCET.
%Indeed, a main result of this paper is a refactorization of a soft function for a jet mass observable which does allow
%for additional resummation in a certain limit (see Section~\ref{sec:highOrder} below).

Now consider non-hemisphere jets.
There has already been some work in SCET on
understanding finite jet-size effects.
For example, jet functions for the $k_T$ and cone jets have been calculated at
NLO~\cite{Jouttenus:2009ns,Ellis:2010rwa} , agreement with
QCD calculations of the 2-jet limit has been shown for small $R$~\cite{Cheung:2009sg},
and RG invariance at the 1-loop level has been demonstrated for jet mass and jet angularity
distributions~\cite{Ellis:2009wj,Ellis:2010rwa}.
These works have provided a number of non-trivial consistency checks on the SCET formalism.
However, it has not yet been demonstrated that any of the jet shapes exponentiate.
Due the presence of multiple scales, such as the jet size $R$ and the amount of in-jet and out-of-jet radiation,
there is plenty of opportunity for logs like $\log(m_L^2/m_R^2)$ in the hemisphere case,
which are not fixed by RGE, to inhibit resummation.

In this paper, we will consider various jet mass observables. We focus on $e^+e^-\to$ dijets events, since analytical
calculations can be performed to order $\alpha_s$ and numerical calculations to $\alpha_s^2$.
To lend concreteness to the discussion, all events are to be clustered using the Cambridge/Aachem (CA) algorithm
(see \cite{Salam:2009jx}).  That is, take the pair of particles with the smallest angle and merge them if the distance
\begin{equation}
 R_{ij} = \frac{1}{2}(1-\cos \theta_{ij})
\end{equation}
is less than $R$. If no particles have $R_{ij}<R$ then stop.
The jets can be ordered by energy, and $E_i, p_i^{\mu}$ will denote the energy and 4-momentum of the $i$-th most
energetic jet.
This parameter $R$ can be translated into
other jet size measures, such as ones used at hadron colliders.

We begin by considering observables which avoid a jet veto. To do this, one has to have an inclusive observable whose
singular region forces the threshold in which both jet masses are small. The example we consider is asymmetric jet masses
and asymmetric thrust, which are defined in Section~\ref{sec:inc}.
In this case, we show that the asymmetric jet mass of the hardest jet does not agree with any calculation in SCET,
while the asymmetric jet mass of the quark (or anti-quark) jet cannot be defined in QCD. When one averages over
both jets, the results do agree, and we check at order $\alpha_s^2$ that the next-to-next-to-leading logarithms
are being resummed.

Next, we move on to observables which have an out-of jet region. To resum the large logarithms of jet mass in this
case, one must reject events in which there is hard radiation in the out-of-jet region. To do this, we introduce
a parameter $\omega$ which controls the out-of-jet energy. Resumming logs of jet mass in this case is trickier than
for asymmetric thrust. We argue that  expanding around a different threshold $R\sim0$, provides a new handle on these complicated multiscale observables.
We postulate a refactorized form of the soft function, valid at small $R$, and then show it gives much improved agreement with the NLO predictions of QCD,
than using SCET alone. This indicates than an expansion around small $R$ maybe be a promising direction for precision jet substructure calculations.

\section{Inclusive Jet Mass: Asymmetric Thrust \label{sec:inc}}
In this section, we consider a simple inclusive jet mass observable, introduced in \cite{Kelley:2010qs}. Cluster the
particles into jets of size $R$ using Cambridge/Aachen (see above). Then asymmetric thrust $\tau_A$ is defined
as the sum of the mass-squared of one primary jet and the mass-squared of everything else in the event. Which jet is chosen to be primary
will be discussed shortly. 

There are three main reasons for considering this odd-sounding observable: First, every particle
contributes to the value of $\tau_A$. So the region with $\tau_A \ll 1$ forces dijet kinematics and the SCET factorization
formula should hold. Second, there is $R$-dependence in the jet mass so one can study $R$-dependence of threshold logs.
This provides a warm up to more complicated exclusive jet mass observables. Third, asymmetric thrust may be useful
to measure jet masses at hadronic colliders. Indeed, it was suggested first in the context of dijet events in $p p$ collisions.
But more simply, in direct-photon\cite{Becher:2008cf} or $W/Z$~\cite{Becher:2011fc} events at high $p_T$, asymmetric thrust would measure the same singular behavior of
the jet mass as any other jet mass measure, but it may be easier to compute and is likely to undergo dynamic threshold
enhancement. The hadron collider applications are not the subject of this paper.

In dijet events at $e^+e^-$ machines, the first question is which jet should be the primary jet for asymmetric thrust. In data (or in full QCD), the obvious candidate
is the hardest jet. Let us call this $\tAo$. That is,
%-----------------------------------------------------------------------------------------
\begin{equation}
 \tAo = \frac{1}{Q^2}(m_1^2 +m_{\bar{1}}^2)
\end{equation}
%-----------------------------------------------------------------------------------------
where $m_1$ is mass of the hardest jet, (largest energy),
and $m_{\bar{1}}$ is the mass of all the particles not clustered into the hardest jet.
A quick study of the kinematics shows that the $\tAo$ distribution in QCD at order
$\alpha_s$ is exactly equal to the $\tau=1-T$ distribution, with no $R$-dependence.
That is,
%-----------------------------------------------------------------------------------------
\begin{multline}
\frac{1}{\sigma_0}\left[ \frac{\rd \sigma}{\rd \tAo}\right]_{\text{QCD}} =\delta(\tAo)
+ \frac{C_F\alpha_s}{4\pi}\left\{
  \left[\frac{-8 \ln \tAo - 6}{\tAo}\right]_+ \right. \\
  \left.
  + \left[  -2 + \frac{2\pi^2}{3} \right]
    \delta(\tAo) +\cdots\right\} \label{tauAqcd}
\end{multline}
%-----------------------------------------------------------------------------------------
where the $\cdots$ are regular in $\tAo$.

Now we turn to the calculation of asymmetric thrust in SCET.  Note that $\tau_A \to 0$ forces dijet kinematics
independent of which jet is chosen to be the primary one.
Thus, the effective theory should agree with QCD in the singular region
for any $R$.
In fact, if we are interested in the
limit that $\tau_A \to 0$, then the size of the jet is irrelevant to the singularities associated with collinear radiation.
For this reason,
we can use the inclusive jet function. Indeed, the use of the inclusive jet function for the hemisphere mass case has to be
justified using the same logic. In contrast, for the soft function, the $R$-dependence is important.  The calculation of the soft function is problematic for $\tAo$
The problem can be traced to the fact that the soft function in SCET
has no access to the energy of the jets. In the region of phase space in which the gluon is soft but not
within $R$ of either quark, the most energetic jet is not well defined. In QCD, there is no such ambiguity.
We conclude that SCET cannot calculate $\tAo$.

Instead,
one might consider calculating $\tAq$, where the primary jet is taken to be the quark.
When the radiation is clustered with the quark, we find that the contribution to the
soft function from radiation inside a cone of size $R$ is
%---------------------------------------------------------------------------------
\begin{multline}
  S^{\in}_R(k,\mu) = \delta(k) + \frac{C_F\alpha_s}{4\pi}
    \left(-2\ln^2\Rf + \frac{\pi^2}{3}\right)\delta(k) \\
+ \frac{C_F\alpha_s}{4\pi} \left[\frac{-16\ln\frac{k}{\mu}+8\ln\Rf}{k}\right]_\star^{[k.\mu]} \label{Sin}
\end{multline}
%---------------------------------------------------------------------------------
where the $\star$-distribution notation can be found in~\cite{Schwartz:2007ib}.
When the radiation is outside the quark's cone, the
contribution from the complimentary region (everything but the cone) is simply
$S^{\in}_{1-R}(k,\mu)$.
When we convolute these two contributions to the soft function with
the inclusive ($R$-independent) jet functions, as in Eq.~\eqref{mmfact}, we
produce the $\tAq$ distribution in SCET. The result is
%---------------------------------------------------------------------------------
\begin{multline}
\frac{1}{\sigma_0}\left[ \frac{\rd \sigma}{\rd \tAq}\right]_\scet =\delta(\tAq)
+ \frac{C_F\alpha_s}{4\pi}\left\{
  \left[\frac{-8 \ln \tAq - 6}{\tAq}\right]_+ \right.\\
\left. + \left[
%\frac{2}{\e^2} +\frac{3}{\e}+
-2+\frac{2\pi^2}{3} - 4 \ln^2\Rf \right]\delta(\tAq) + \cdots \right\} \label{tauq}
\end{multline}
%---------------------------------------------------------------------------------
Comparing to Eq.~\eqref{tauAqcd} we see
that the asymmetric thrust distribution in SCET has $R$ dependence, while the QCD result for $\tAo$ does not,
 so the distributions do not agree. Of course, it is not unreasonable that they do not agree, since they are different observables.

If we repeat the QCD calculation, but always defining the primary jet as the quark jet, we find
a result which agrees exactly with the SCET prediction, Eq.~\eqref{tauq}, up to non-singular terms.
Unfortunately, this quark-jet asymmetric thrust is not infrared safe, and therefore cannot be used to measure jet mass.
Indeed, testing $\tAq$ at next-to-leading order by taking the primary jet to be the hardest one containing a quark parton, we find negative cross sections.

Instead of attempting to define a quark jet in QCD (see~\cite{Banfi:2006hf} for an example of how this might be done), we can
get agreement by averaging the jets. Note first that the hardest jet asymmetric thrust, which
we calculated in QCD, is not the average of the quark and anti-quark asymmetric thrust in SCET. However, if we calculate
asymmetric thrust in QCD with the second hardest jet as primary, we find
%----------------------------------------------------------------------------------------------------
\begin{multline}
\frac{1}{\sigma_0}\left[ \frac{\rd \sigma}{\rd \tAt}\right]_{\qcd} =\delta(\tAt)
+ \frac{C_F\alpha_s}{4\pi}\left\{ \left[\frac{-8 \ln \tAt - 6}{\tAt}\right]_+ \right. \\
\left. + \left[
%\frac{2}{\e^2} +\frac{3}{\e}+
-2+\frac{2\pi^2}{3}  - 8 \ln^2\Rf\right]\delta(\tAt) +\cdots\right\} \label{tauAtqcd}
\end{multline}
%----------------------------------------------------------------------------------------------------
Thus the average of the harder and softer jet asymmetric thrust distributions in QCD
is the same as the average of the quark and anti-quark distributions in SCET at order $\alpha_s$. While this
may not be surprising, it illustrates the importance of being careful not to discuss quark jet and gluon-jet masses in
QCD without a proper infrared safe observable definition~\cite{Gallicchio:2011xc,Gallicchio:2011xq}.

%----------------------------------------------------------------------------------------------------
\begin{figure}[t]
\includegraphics[width=0.95\hsize]{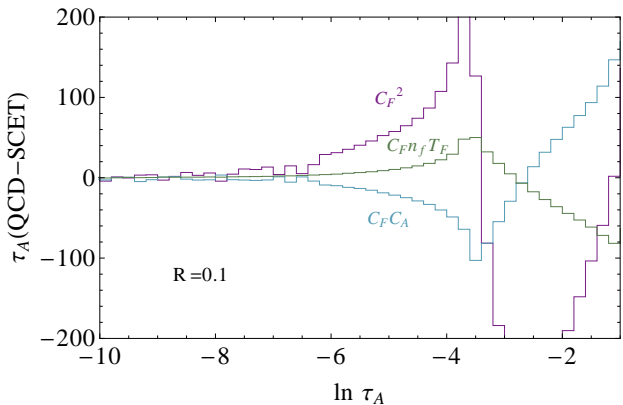}
\caption{Difference between QCD and SCET for the sum $\tau_{A_1}\frac{\rd\sigma}{\rd \tau_{A_1}}+ \tau_{A_2}\frac{\rd\sigma}{\rd \tau_{A_2}}$ at order $\alpha_s^2$. That the difference
vanishes at very small $\tau_A$ (left side), shows that SCET reproduces all of the large logs of this distribution up to NNLL.
\label{fig:tauA}
}
\end{figure}
%----------------------------------------------------------------------------------------------------

Now that the observables agree at order $\alpha_s$ we can check resummation by expanding the resummed results in SCET
to order $\alpha_s^2$ and comparing against the output of the numerical QCD calculation using the program {\sc event 2}~\cite{Catani:1992ua}.
We look at the observable
\begin{equation}
 \frac{\rd \sigma}{\rd \tau_A}
 = \frac{\rd \sigma}{\rd \tAo}
+\frac{\rd \sigma}{\rd \tAt}
\end{equation}
and compare to the result of adding the quark and anti-quark jet asymmetric thrust distributions in SCET.

\begin{figure}[t!]
\includegraphics[width=0.95\hsize]{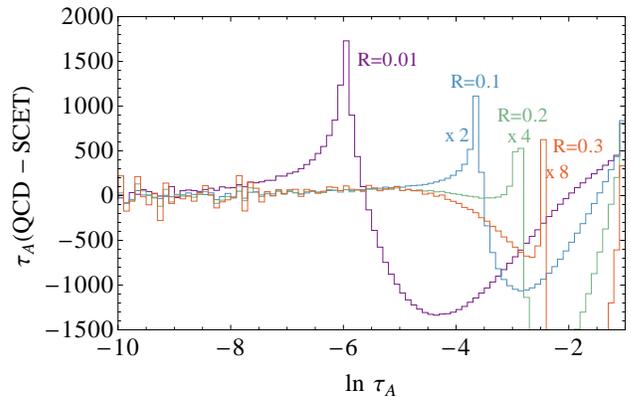}
\caption{Difference between QCD and SCET for $\tAo \frac{\rd\sigma}{\rd \tAo}$ at order $\alpha_s^2$ for a variety of jet sizes.
\label{fig:tauArs}
}
\end{figure}

Figure~\ref{fig:tauA} shows the difference between the QCD and SCET predictions for $\tau_A$ as a function of $\ln \tau_A$ in the singular region, separated
by color structure. Figure~\ref{fig:tauArs} shows the distribution for a number of different values of $R$. In all cases, there is excellent convergence at $\tau_A \to 0$,
showing that SCET accounts for all the singularities at NLO. This demonstrates that SCET can resum all of the large logarithms of $\tau_A$ for any $R$. At very small $R$,
these may not be the dominant part of the distribution, due to large $\ln R$ terms, but we have not attempted to resum logs of $R$ for $\tau_A$.

As observed in~\cite{Kelley:2010qs}, $\tau_A$ may be a promising to measure the jet mass in $W/Z/\gamma$+jet events at hadron colliders. It can be calculated
at the machine threshold and should undergo dynamical threshold enhancement making it relevant at the partonic threshold as well. An alternative to invoking
 dynamical threshold enhancement would be to have a jet mass observable which is not sensitive to the whole event. For example, one could look at the mass of a jet
in exclusively 2-jet events. To do this requires a veto on additional radiation. So we now turn to the calculation of jet mass with a jet veto in SCET.

%----------------------------------------------------------------------------------------------------
%----------------------------------------------------------------------------------------------------

\section{Exclusive Jet Mass: Dijets with a Veto}
In this section, we will consider the exclusive case, where jets are clustered using Cambridge/Aachen, as before,
but only the two hardest jets are considered. For asymmetric thrust, there was one primary jet, with the secondary
jet containing all the other radiation in the event. For an exclusive dijet sample, we have to veto additional jets.
We do this by imposing an energy cutoff $\omega$ on the 3rd most energetic jet: $E_3 < \omega$ (at tree level and $\mathcal{O}(\alpha_s)$ this is equivalent
to vetoing on the total out-of-jet radiation, up to a contribution in which the jet mass is exactly zero).
We will consider several different dijet mass observables and explore to what extent SCET can reproduce the QCD results
and perform resummation of the large logarithms.

The first observable we will consider is the doubly differential cross section
in the masses $m_1$ and $m_2$ of the two most energetic jets.
As we saw with asymmetric thrust, one cannot order the jets by energy in SCET
since, after matching, the operator only has access to the jets' energies through the labels, each of
which has the same magnitude. Instead in SCET one can calculate the quark and anti-quark jet mass, but these
are not infrared safe in QCD. As with asymmetric thrust, we will see that a meaningful comparison
can be made between an average over the two jets or more simply the sum of the jet masses squared.

First consider the QCD calculation of $\rd^2 \sigma/ \rd m_{1}^2 \rd m_{2}^2$.
The tree level process $e^+ e^- \to q \bar{q}$ and its one-loop virtual corrections
contribute to the $\delta(m_1^2) \delta(m_2^2)$ since the final state partons are massless.
This contribution is given by
%----------------------------------------------------------------------------------------------------
\begin{align}
&\frac{1}{\sigma_0} \left[ \frac{\rd^2 \sigma}{\rd m_1^2 \rd m_2^2}\right]^{\text{ tree + virtual} }
=
  \delta(m_1^2) \delta(m_2^2)
\nn\\
&\qquad \times
  \left\{
   1 +
  \frac{\alpha}{4\pi} C_F
  \left(
  -\frac{4}{\e^2}
  -\frac{6}{\e }
  -16
  +\frac{7\pi ^2}{3}
  \right)
  \right\}
\end{align}
%----------------------------------------------------------------------------------------------------
For the real emission process, $e^+ e^- \to q \bar{q} g$, there are non trivial contributions to the jet mass
distribution.

\begin{figure}[t]
\begin{center}
\includegraphics[width=6cm]{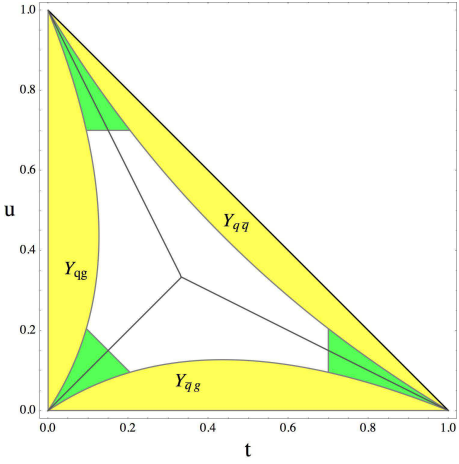}
\caption{The allowed kinematic region for $e^+ e^- \to q \bar{q} g$ in variables $u, t$.  In the yellow regions,
$Y_{ij}$, the Cambridge/Aachem requirements have been met and partons $i,j$ have been clustered. The green region
corresponds to $E_3 < \omega$.  In this plot, $R = 0.4$ and $\omega = 0.15 Q$. }
\label{PS1}
\end{center}
\end{figure}

The differential cross section for real emission of a gluon
in $d = 4 - 2\e$ dimensions, leading order in $\alpha_s$, is given by
%----------------------------------------------------------------------------------------------------
\begin{multline}
\frac{1}{\sigma_0}\frac{\rd^2 \sigma}{\rd u \rd t}
 = \frac{\alpha_s}{4\pi} C_F \frac{2}{ t^{1 + \e} u^{ 1 + \e } (1 -t-u)^{\e } }   \nn\\
\qquad \times \left( (1-t)^2 + (1-u)^2 -\e (t+u)^2 \right).
\end{multline}
%----------------------------------------------------------------------------------------------------
where
%----------------------------------------------------------------------------------------------------
\begin{align}
s &= (p_q + p_{\bar{q} } )
&t &= (p_q + p_{g } )
&u &= (p_{\bar{q} } + p_g ).
\end{align}
%----------------------------------------------------------------------------------------------------
The kinematically allowed regions are shown in Fig.~(\ref{PS1}).
%----------------------------------------------------------------------------------------------------
%----------------------------------------------------------------------------------------------------
The yellow regions shown in the figure correspond the regions in phase space where the C/A clustering requirements
are met.  These regions are labeled $Y_{ij}$ for $i,j = q, \bar{q}, g$, where the subscript corresponds to
to the pair of partons that have been clustered. In these regions, the clustered jet has the most energy and so
we assign $p_1^{\mu} = p_i^{\mu} + p_j^{\mu}$, whereas $p_2^{\mu}$ is assigned to the momentum of the remaining parton.
The shaded green region accounts for the case when the partons are not clustered so there are three massless jets;
however, these are still counted as dijet events since $E_3 < \omega$ and thus contribute to the
$\delta(m_1^2) \delta(m_2^2)$ part of the distribution.  After performing the integrations,
the real emission contribution is given by
%----------------------------------------------------------------------------------------------------
\begin{multline}
\frac{1}{\sigma_0} \left[ \frac{\rd^2 \sigma}{\rd m_1^2 \rd m_2^2}\right]^{\text{ real} }
  =
  \delta(m_1^2) \delta(m_2^2)
\nn\\
\qquad +
  \frac{\alpha}{4\pi} C_F
  \delta(m_2^2)
  \left\{
  \left(
  \frac{4}{\e^2}
  +\frac{6}{\e }
  +14
  -\frac{5\pi ^2}{3}
  \right.
  \right.
\nn \\
 \qquad
\left.
\left.
-8\log \frac{R}{1-R} \log \frac{2\omega}{Q}
+f_{\omega}(R)
\right)
\delta(m_1^2)
\right.
\nn \\
\qquad
  +
  \left.
  \left[
  \frac{  -6 + 8 \log \frac{R}{1-R} - 8\log \frac{m_1^2}{Q^2} }{m_1^2}
  \right]_{\ast}
+ \cdots
\right\}
\end{multline}
%----------------------------------------------------------------------------------------------------
where $\cdots$ denotes terms at higher order in $m_1^2$ and
%----------------------------------------------------------------------------------------------------
\begin{align}
&
f_{\omega}(R)
=
-4 \log R \log \frac{R}{1-R}
-4 \text{Li}_2(1 - R)
+4 \text{Li}_2(R)
\nn \\
&\qquad
+\frac{8\omega}{Q}\left( 2 + \frac{3}{2} \log \frac{R}{1-R} \right)
+ \mathcal{O}\left( \frac{\omega^2}{Q^2} \right)
\label{fomega}
\end{align}
%----------------------------------------------------------------------------------------------------
The result of adding the tree, virtual, and real emission contributions is
%----------------------------------------------------------------------------------------------------
\begin{multline}
\frac{1}{\sigma_0}\left[ \frac{\rd^2 \sigma}{\rd m_1^2 \rd m_2^2}\right]_\qcd
  =
  \delta(m_1^2) \delta(m_2^2) +
  \frac{\alpha}{4\pi} C_F
  \delta(m_2^2)\\
  \times \left\{
  \left(
  -2
  +\frac{2\pi ^2}{3}
  \right.
  \right.
\left.
\left.
-8\log \frac{R}{1-R} \log \frac{2\omega}{Q}
+f_{\omega}(R)
\right)
\delta(m_1^2)
\right.
 \\
\qquad
  +
  \left.
  \left[
  \frac{  -6 + 8 \log \frac{R}{1-R} - 8\log \frac{m_1^2}{Q^2} }{m_1^2}
  \right]_{\ast}
+ \cdots
 \right\} \,.
 \label{toQCD}
\end{multline}
%----------------------------------------------------------------------------------------------------

As we saw in the last section, SCET does not have access to energy ordering of the jets,
and the closest observable we can calculate is $\rd^2 \sigma/\rd m_q^2 \rd m_{\bar{q}}^2$,
which we turn to now.
The derivation of the factorization theorem for $\rd^2 \sigma/\rd m_q^2 \rd m_{\bar{q}}^2$
is almost identical to the derivation of the factorization theorem for thrust.
For very small $m_{q,\bar{q}}$, the jet size $R$ affects the collinear sector only at
subleading power of $m_{q,\bar{q}}$. Therefore, as in the hemisphere
mass case, we can use the inclusive ($R$-independent) jet function.
The factorization formula then reads
%----------------------------------------------------------------------------------------------------
\begin{multline}
 \frac{1}{\sigma_0} \frac{\rd^2 \sigma}{\rd m_q^2 \rd m_{\bar{q}}^2} =H(Q^2,\mu) \int \rd k_q \rd k_{\bar{q}} \\
\times  J(m_q^2 - k_q Q) J(m_{\bar{q}}^2 - k_{\bar{q}} Q) S_R(k_q, k_{\bar{q}},\omega, \mu)
\label{SCETfact} \,.
\end{multline}
%----------------------------------------------------------------------------------------------------
This soft function is defined as (see~\cite{Fleming:2007qr})
%----------------------------------------------------------------------------------------------------
\begin{multline}
  S_R(k_q,k_{\bar{q}},\omega,\mu) =\sum_{X_s} \langle 0 | Y_{\bar{n}}^\dagger Y_n | X_s \rangle \langle X_s |  Y_{n}^\dagger Y_{\bar{n}} | 0\rangle \\
\times \delta( k_{q} - n\cdt p_X^q) \delta(k_{\bar{q}} - \bar{n} \cdt p_X^{\bar{q}}) \theta(\omega-k_0^\out) \,,
\end{multline}
%----------------------------------------------------------------------------------------------------
where $n^\mu$ and $\bar{n}^\mu$ are lightlike 4-vectors in the $q$ and $\bar{q}$ jet directions, respectively,
$Y_n$ are soft Wilson lines,
$X_s$ is the soft radiation,
 $p_X^q (p_X^{\bar{q}})$ the component of its momenta in the quark (anti-quark) jet, and $k_0^\out$
is the energy the hardest jet of size $R$ in the inter-jet region.

At order $\alpha_s$, the soft function can be written as a product of two in-cone soft functions
and an out-of-cone soft function: $S_R=S^\in_R(k_q,\mu)S^\in_R(k_{\bar{q}},\mu) S^\out_R(\omega,\mu)$.
The in-cone soft function is the same as for asymmetric thrust and given in Eq.~\eqref{Sin}.
The out-of-cone soft function is
%----------------------------------------------------------------------------------------------------
\begin{multline}
  S^\out_R(\omega,\mu) = 1 +
\frac{C_F\alpha_s}{4\pi}\Big[ -8 \ln \Rf \ln\frac{2\omega}{\mu} \\
+2\ln^2\Rf+ f_0(R)\Big] \,,
\end{multline}
%----------------------------------------------------------------------------------------------------
with $f_0(R)$ given in Eq.\eqref{fomega}.  These results agree with limits of expressions given in~\cite{Ellis:2010rwa}.

Combining this soft function with the hard and inclusive jet functions, we get
%----------------------------------------------------------------------------------------------------
\begin{multline}
\frac{1}{\sigma_0} \left[\frac{\rd^2 \sigma}{\rd m_q^2 \rd m_{\bar{q}}^2}\right]_\scet
  =
  \delta(m_q^2) \delta(m_{\bar{q}}^2) +
  \frac{\alpha_s}{4\pi }C_F \Biggl\{\\ \left(-2
 +\frac{2\pi ^2}{3}
 \right.
\left.
-8\log \frac{R}{1-R} \log \frac{2\omega}{Q}
+f_{0}(R)
\right)
\delta(m_q^2) \delta(m_{\bar{q}}^2)
 \\
\qquad
  +
  \left[
  \frac{  -6 + 8 \log \frac{R}{1-R} - 8\log \frac{m_q^2}{Q^2} }{2m_q^2}
  \right]_{\ast} \delta(m_{\bar{q}}^2)
\\
\qquad
  +
  \left[
  \frac{  -6 + 8 \log \frac{R}{1-R} - 8\log \frac{m_{\bar{q}}^2}{Q^2} }{2m_{\bar{q}}^2}
  \right]_{\ast} \delta(m_{q}^2)
+ \cdots
\Biggr\}
\end{multline}
%----------------------------------------------------------------------------------------------------

We can now compare SCET directly to the doubly differential distribution of the
harder and softer jet mass, which is the natural quantity to measure experimentally, and we have
calculated in QCD.
The coefficient of $\delta(m_q^2) \delta(m_{\bar{q}}^2)$
in the SCET distribution matches the coefficient of $\delta(m_1^2) \delta(m_2^2)$ from  Eq.\eqref{toQCD}
with $f_0(R)$ appearing instead of $f_\omega(R)$; however, the remaining distribution is symmetric in
$m_q \leftrightarrow m_{\bar{q}}$ in SCET, but not symmetric in $m_1 \leftrightarrow m_2$ for QCD. So the doubly differential distributions are
different. We anticipated this, but it is still instructive to see that the mass of the hardest jet is
not simply related to any projection of the doubly differential distribution in SCET.

Although the hardest jet mass is not reproduced in SCET, one can compute the average of the
distributions of the two jet masses. If one integrates one jet mass up to the full kinematic limit,
in general there will be non-global logarithms. The result is like comparing the left-hemisphere
mass in QCD and in SCET, which do not agree due to non-global logarithms due to extra hard emissions
within one jet.  However, at small $R$, there is a natural cutoff $m < Q^2 R$. Then we can compute
%----------------------------------------------------------------------------------------------------
\begin{align}
\left[\frac{\rd \sigma}{\rd m^2}\right]_\text{QCD}
\!\!
& =
\int_{0}^{Q^2 R} \rd m_1^2 \int_{0}^{Q^2 R} \rd m_2^2 \ \frac{\rd^2 \sigma}{\rd m_{1}^2 \rd m_{2}^2}
\nn \\
& \qquad
  \times \frac{1}{2}
  \bigl[
  \delta( m^2 - m_1^2 ) + \delta( m^2 - m_2^2 )
  \bigl] \, .
\end{align}
%----------------------------------------------------------------------------------------------------
This can be compared to the analogous expression in SCET, using $m_q$ and $m_\qbar$ instead of $m_1$ and $m_2$.
The two distributions agree, as they should.

Rather than look at the average jet mass distribution, we can instead look at a thrust-like variable,
the sum of the squares of the jet masses. We call this \textbf{jet thrust}, and denote it by $\tau_\omega$.
By definition, $\tau_{\omega}  = (m_1^2 + m_2^2)/Q^2$ where events with $E_3 < \omega$ have been vetoed.
At leading order,
$\rd \sigma / \rd \tau_{\omega}$ can be calculated by integrating the doubly differential distribution computed above.
%----------------------------------------------------------------------------------------------------
\begin{align}
&
\left[\frac{\rd \sigma}{\rd \tau_\omega}\right]_\text{QCD}
  =
  \int_{0}^{Q^2R} \rd m_1^2 \int_{0}^{Q^2R}\rd m_2^2 \ \frac{\rd^2 \sigma}{\rd m_{1}^2 \rd m_{2}^2}
\nn \\
&
  \qquad
  \times \delta( \tau_{\omega} - m_1^2/Q^2 - m_2^2/Q^2)
\end{align}
%----------------------------------------------------------------------------------------------------
The terms singular in $\tau_{\omega}$ in full QCD at order $\alpha_s$, are
%----------------------------------------------------------------------------------------------------
\begin{multline}
\frac{1}{\sigma_0}\left[\frac{\rd \sigma}{\rd \tau_\omega}\right]_\qcd =
 \delta(\tau_\omega) + \frac{C_F \alpha_s}{4\pi} \left[ \frac{-8\ln \tau -6 + 8\ln\Rf}{\tau_\omega}\right]_+ \\
+ \frac{C_F \alpha_s}{4\pi}\left[ -2 + \frac{2\pi^2}{3} - 8\ln\Rf\ln\frac{2\omega}{Q}
+f_\omega(R)\right] \delta\left(\tau_\omega\right).
\label{tomQCD}
\end{multline}
%----------------------------------------------------------------------------------------------------
When $\tau_{\omega} \ll 1$, each jet mass is forced to be small and so SCET is valid in this regime.
In the effective theory, we take $\tau_{\omega} = (m_q^2 + m_{\bar{q}}^2)/Q^2$ and its distribution
can be obtained via
%----------------------------------------------------------------------------------------------------
\begin{align}
&
\left[\frac{\rd \sigma}{\rd \tau_\omega}\right]_\scet
  =
  \int_{0}^{Q^2R} \rd m_q^2 \int_{0}^{Q^2R}\rd m_{\bar{q}}^2 \  \frac{\rd^2 \sigma}{\rd m_{q}^2 \rd m_{\bar{q}}^2}
\nn \\
& \qquad
  \times \delta( \tau_{\omega} - m_q^2/Q^2 - m_{\bar{q}}^2/Q^2)
\end{align}
%----------------------------------------------------------------------------------------------------
The terms singular in $\tau_{\omega}$ in SCET at order $\alpha_s$, are
%----------------------------------------------------------------------------------------------------
\begin{multline}
\frac{1}{\sigma_0}\left[\frac{\rd \sigma}{\rd \tau_\omega}\right]_\scet =
 \delta(\tau_\omega) + \frac{C_F \alpha_s}{4\pi} \left[ \frac{-8\ln \tau_{\omega} -6 + 8\ln\Rf}{\tau_\omega}\right]_+ \\
+ \frac{C_F \alpha_s}{4\pi}\left[
  -2
  +\frac{2\pi ^2}{3}
  - 8\ln\Rf\ln\frac{2\omega}{Q}
  +f_0(R)\right] \delta\left(\tau_\omega\right).
\label{tomSCET}
\end{multline}
%----------------------------------------------------------------------------------------------------
This result is the same as the QCD result with $f_{0}(R)$ appearing instead of $f_{\omega}(R)$.
Thus at LO, SCET reproduces the parts of QCD
which are singular in $\tau_\omega$, up to power corrections in $\omega/Q$, but including all of the $R$ dependence.

\section{Predicting higher order terms}
\label{sec:highOrder}
In the previous sections we explored doubly and singly differential jet mass distributions in SCET and in QCD.
For some observables, we saw that SCET reproduces the singular behavior of QCD at order $\alpha_s$. In particular,
for jet thrust, $\tau_\omega$, which we defined as the sum of the jet masses squared normalized to the machine energy $Q$
with a energy veto $E < \omega$ on the third hardest jet, we found agreement with QCD for any $R$ at order $\alpha_s$, up to power corrections
in $\omega/Q$.
Next, we would like to know to what extent this agreement persists to higher order.
The question can be phrased succinctly as asking which of the coefficients $C_{ij}$ in the expansion
\begin{equation}
 \tom \frac{\rd \sigma}{\rd \tom} = \sum C_{ij}(R,\omega) \alpha_s^i \ln^j \tom
\label{eq:Cij}
\end{equation}
can be predicted in SCET.

There are two parts to this question. First, if one were able to do calculations to all order in SCET,
would the coefficients come out correct?  As long as the entire
singular region is described by soft and collinear degrees of freedom, the answer should be yes.
This is true for $\tom$ as long as $\omega \ll Q$  for any $R$. So SCET should get all the $C_{ij}(R,\omega)$ correct
up to power corrections in $\omega/Q$.  This can be seen already at order $\alpha_s$ in Eq.~\eqref{tomSCET} where the
SCET distribution agrees with the QCD distribution in Eq.~\eqref{tomQCD} exactly in $R$ when terms of order $\omega/Q$ or smaller
are dropped.

The second part is whether the coefficients $C_{ij}(R,\omega)$ with $i>1$ can be predicted using factorization. That is, which of these coefficients
can we get without actually doing the explicit loop integrals? For example, non-global logarithms appear in the doubly differential hemisphere
mass distribution. These are reproduced in SCET~\cite{Kelley:2011ng,Hornig:2011iu}, but it is so far unclear how to reproduce them without doing calculations
to all orders in perturbation theory. We would like to know what can and cannot be predicted, in what limits. We would also like to know
to what extent terms which we cannot predict or calculate are actually important.

Since the anomalous dimensions of the jet and hard functions are known to 3-loops, the $\mu$-independence of the $\tau_\omega$ distribution
determines the anomalous dimension of the soft function to 3-loops as well.
Note that since the jet and hard functions are completely independent of $R$ and $\omega$, none of the anomalous dimensions have any dependence on $R$ or $\omega$ at all.
With the finite part of the 1-loop soft function known, this should
naively allow us to resum the next-to-next-to-leading logs (NNLL). That would give us all the terms down to $1/\tau_\omega$ in the $\alpha_s^2$ distribution.
This is certainly the case for hemispheres ($R=\frac{1}{2}$). However, for smaller $R$ these anomalous dimensions are not enough to predict the NNLL,
or even NLL distribution.

The failure of SCET to describe $\tau_\omega$ at NLL can be seen already in Eq.\eqref{tomSCET}: at NLL, the 1-loop anomalous dimensions should determine the $\alpha_s /\tau_\omega$ terms.
But since the jet and hard anomalous dimensions lack $R$ dependence, there is no way they can predict the $\alpha_s \ln\Rf/ \tau_\omega$ piece.
Thus, without insights into the soft function beyond the original factorization, SCET cannot even resum $\tau_\omega$ to NLL.

\begin{figure}[t!]
\includegraphics[width=0.95\hsize]{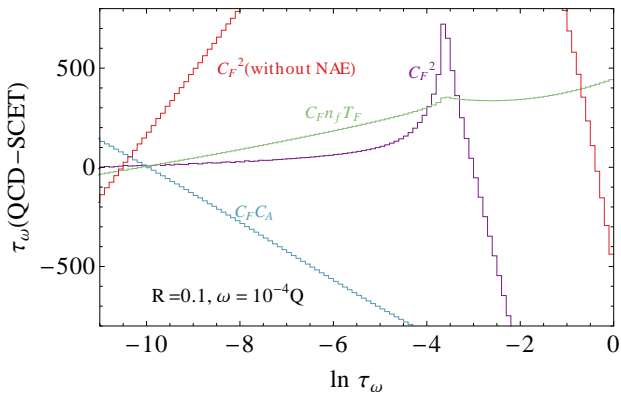}
\caption{Difference between QCD and SCET for the $(\alpha_s/4\pi)^2$ coefficient of $\tau_\omega \frac{\rd \sigma}{\rd \tau_\omega}$ for the different color structures.
After using non-Abelian exponentiation (NAE) the $C_F$ curve should go to zero at small $\tau_\omega$, up to power corrections in $\omega$.
}
\label{fig:diffplotNAE}
\vspace{-0.4cm}

\end{figure}

To proceed, we first of all observe that the troublesome term has a coefficient $C_F$.
Due to the non-Abelian exponentiation (NAE) theorems~\cite{Frenkel:1984pz,Gatheral:1983cz} the $\alpha_s^n C_F^n$ terms
in the soft function are determined exactly by the 1-loop result. Thus we can resum at least this much $R$-dependence by simply exponentiating
the NLO soft function (in position space). Doing so predicts an NLL term $C_F^2 \alpha_s^2 \ln^2 \tau/\tau$, as well as $\alpha_s^2 \ln \tom/\tom$ and
$\alpha_s^2/\tom$ NNLL terms.

To see how well SCET is doing, we will explore the order $\alpha_s^2$ distribution of $\tau_\omega$. This can be calculated in QCD
numerically using the program {\sc event 2}~\cite{Catani:1996jh}.
 In Figure~\ref{fig:diffplotNAE} we show
difference between QCD and SCET predictions
at order $\alpha_s^2$ for the different color structures.
The $C_F^2$ curve before and after using non-Abelian exponentiation  are shown.
We see that after NAE, the $C_F^2$ curve goes to zero as $\tom \to 0$, while the $C_F C_A$ and $C_F n_f T_F$ color structures are clearly missing $\ln \tom Q/\omega$ pieces, which affects the slope of these plots.
Note that the curves cross at roughly $\tom \sim \omega=10^{-4} \sim e^{-9}$, indicating a missing $\ln\omega/\tom$ piece.
The approach of the $C_F$ term to zero gives us a benchmark for where we should expect the curves to vanish, and how large the $\omega/Q$ power
corrections are.

 \subsection{Refactorization}
To further progress, we will now argue that when
%-----------------------------------------------------------------------------------
\begin{equation}
 \omega \lesssim k_L, k_R \qquad \text{and} \qquad R \ll 1
\end{equation}
%-----------------------------------------------------------------------------------
the soft function factorizes into four parts
%-----------------------------------------------------------------------------------
\begin{multline}
  S_{R}(k_L,k_R,\omega,\mu) =
  S_{R}^{\in}(k_L,\mu)
  S_{R}^{\in}(k_R,\mu)
  S_{R}^\out(\omega,\mu) \\
\otimes S_f(k_L,k_R,\omega)
 \label{factsoft} \,,
\end{multline}
%-----------------------------------------------------------------------------------
Here, $\otimes$ indicates a convolution. Similar refactorization formulas have appeared before~\cite{Ellis:2009wj,Ellis:2010rwa} (and after~\cite{Hornig:2011tg}), but of a more general nature.
Without specifying the regime in which the refactorization is valid, and more importantly, properties of the objects appearing in the refactorization,
there is no content in Eq.~\eqref{factsoft}.

To see that Eq.~\eqref{factsoft} by itself is contentless, recall that the hemisphere soft function can be written as
%-----------------------------------------------------------------------------------
\begin{multline}
  S_{\text{hemi}}(k_L,k_R,\mu) =
  \Pi(k_L,\mu) \Pi(k_R,\mu) \otimes S_f(k_L,k_R) \, ,
\end{multline}
%-----------------------------------------------------------------------------------
where $\Pi(k,\mu)$ are the soft RG factors which compensate for the RG evolution of the jet and hard functions. This form is forced by
the hemisphere factorization formula in SCET~\cite{Fleming:2007qr,Schwartz:2007ib,Hoang:2008fs,Chien:2010kc}. But $S_{\text{hemi}}=S_{R=\frac{1}{2}}$,
so we can find a trivial solution for~\eqref{factsoft} by setting $S_R^{in}(k,\mu) = \Pi(k,\mu)$ and $S_R^\out = 1$.

We propose that $S_R^\in$ and $S_R^\out$ can be thought of as soft functions in their own right, with their own anomalous dimensions. Only $S_R^\in$ has a $\log \mu$ in its
anomalous dimension, since the soft-collinear singularities are within the jets. We conjecture that the regular anomalous dimensions split as
\begin{align}
\gamma^\out_{S} &= -\gcusp \ln\Rf  + \gamma_\Delta \label{eq:outin}\\
\gamma^\in_{S} &= \gcusp \ln\Rf + \gamma_S - \gamma_\Delta \nonumber
\end{align}
where $\gcusp$ is the cusp anomalous dimension. In particular, all of the $R$-dependence in the anomalous dimensions in proportional to $\gcusp$.

We will argue that $\gamma_\Delta=0$ as well, which is consistent with the $R=\frac{1}{2}$ limit, although since our refactorization argument is
valid at small $R$, this is not really justified.
We also set $S_f(k_L,k_R,\omega) = 1$, which is to say we ignore non-global logs. We will see that the refactorization is still in good agreement
with full QCD independent of the non-global log issue.
With this conjecture, the refactorization allows us expand around small $R$
to predict non-trivial parts of the $\tom$ distribution, for $\omega/Q \lesssim \tom \ll R \ll 1$,
which we will compare our prediction to the fixed order result.

A complete proof of refactorization (or more precisely, why $S_R^\out(\omega,\mu)\ne 1$) would ideally involve
operator definitions of the components $S_R^\in$ and $S_R^\out$. That way we can ascribe anomalous dimensions to the components, which can,
at least in principle, be calculated order-by-order in perturbation theory. However, it may not be possible, or even necessary, to have operator
definitions for refactorization. For example, in the hemisphere case, we do not have operator definitions of the components of $S_{\text{hemi}}$. Instead,
$\Pi(k,\mu)$ is defined to compensate the hard and jet function evolution, and $S_f$ is defined as whatever is left over. We aim at this point for
the more modest goal of having a similar algorithmic definition of the components of $S_R$, leaving a more complete non-perturbative understanding for
future work.

In Section~\ref{sub:ref}, we give a heuristic argument for why the soft function should refactorize at small $R$ (and $\omega \lesssim k_L, k_R$).
We 
%discuss a conjecture for the form of $S_R^\out$ and 
present phenomenological evidence in support of this refactorization conjecture in Section~\ref{sub:pred}.

\subsection{Heuristic argument for refactorization \label{sub:ref}}
To justify the refactorization, we will show that the kinematic restriction small $R$ and $\omega \lesssim k_L, k_R$ allows us to apply
soft-collinear factorization~\cite{Bauer:2001yt}. The argument we present here is similar to one used in~\cite{ninja} for when two subjects become collinear
inside a larger jet.

To begin the argument, let us define the ``left''
jet to be in the  $n^\mu$ direction and the ``right'' jet to be in the $\nbar^\mu$ direction.
A particle within the left jet is kinematically restricted to have
$k_L = k^+ < \Rf \: k^- $, where $k^- = \nbar \cdot k$ and $k^+  = n \cdot k$. With the on-shell condition, the transverse
momentum then scales like $k_{\perp} \sim \sqrt{ k^+ k^-}
\sim \sqrt{\Rf} k^- $.
%The right jet is restricted analogously.
So, when $R\ll1$, the jet is forced to have collinear scaling
$(k^- , k^+ , k_{\perp} ) \sim \frac{k_L}{R} (1,R,\sqrt{R})$.  Radiation in the right jet is restricted analogously.
Outside the jet, the soft momentum is only required to have $k_\out \sim k_L (1,1,1)$.
Thus, there
are three sectors, $|X_\out\rangle$, which has ultra-soft scaling and $|X^L_\in\rangle$ and $|X^R_\in \rangle$ which
are soft, with respect to the original collinear momentum, but collinear with respect to the soft radiation outside the jets.
We say these have soft-collinear scaling.
The relevance of soft-collinear modes was also noticed in~\cite{ninja} in a different context.

%
%
% We are just interested in the properties of matrix elements of Wilson lines in a
% restricted phase space $|X_s\rangle = |X_{\in}^{L}\   X^{R}_{\in} \  X_\out \rangle$,
% where $| X^{L}_{\in} \rangle$ and $| X^{R}_{\in} \rangle$ are the radiation in the left and right handed jets,
% respectively, and $|X_\out \rangle$ is the radiation outside the jets.
%
% Our requirement $\omega \lesssim k_L, k_R$ implies that the radiation outside the jets scales like the small
% component of the radiation in the jets $k_{L,R}(1,1,1)$.  Modes with this scaling property will be called ultra-soft.
% Radiation in the jet can have both ultra-soft and soft-collinear scaling, however radiation outside
% is restricted to ultra-soft scaling only.
%
%
%The soft Hilbert space can be written in terms of the soft-collinear and ultra-soft modes:
%%-----------------------------------------------------------------------------------
%\begin{align}
%& | X_L^{\in} \rangle  =  | X_{n c}^{\in}  X_{\rm us}^{L, \in} \rangle  \\ \nn
%& | X_R^{\in} \rangle  =  | X_{\nbar c}^{\in} X_{\rm us}^{R, \in} \rangle  \\ \nn
%& | X^{\out}  \rangle  =  | X_{\rm us}^{\out} \rangle
%\end{align}
%%-----------------------------------------------------------------------------------
%Here $| X_{n, \nbar c}^{\in} \rangle$ is soft-collinear radiation in the $n$ or $\nbar$ direction
%restricted to be in the jet, $| X_{\rm us}^{L,R, \in} \rangle$ is ultra-soft radiation in the $L,R$ jet and
%$| X_{\rm us}^{L,R, \in} \rangle$ is the ultra-soft radiation outside of any jet.
%

It is interesting to observe the soft-collinear modes are formally harder than the original soft modes: in terms of
$\tom\sim k_L/Q$, the soft collinear modes scale like $Q\tom(\frac{1}{R},1,\frac{1}{\sqrt{R}})$ while the
original soft modes scaled like $Q \tom (1,1,1)$. However, the soft-collinear modes are still parametrically
softer than the original collinear modes, which scale like $Q\tom (\frac{1}{\tom},1,\frac{1}{\sqrt{\tom}})$,
since $\tom \ll R$. Thus, both the soft-collinear and ultrasoft modes have eikonal interactions with the original Wilson lines,
and hence are present in the original soft function.

Now consider the original Wilson line
\begin{equation}
Y_{n}( x )
  = {\rm P} \exp
    \left[ ig \int_0^{\infty} ds\  n \cdot A(x + sn ) \right].
\label{Ydef}
\end{equation}
The gauge fields in this Wilson line includes anything with Eikonal interactions with the original collinear fields,
which includes both the ultra-soft and soft-collinear modes. Thus, we can write
$A_\mu = A_\mu^{n {\text s}} + A_\mu^{\bar{n} {\text s}} + A_\mu^{\text{us}}$.
The soft-collinear modes and ultra-soft modes interact with each other the same way that ordinary collinear and soft modes interact.
 In any situation in which
modes have collinear and ultra-soft scaling, the interactions between the soft and collinear modes
can be removed from the Lagrangian through a BPS
field redefinition~\cite{Bauer:2001yt}:
%-----------------------------------------------------------------------------------
\begin{equation}
 A_{ns}      \to ( Y_n^{\rm us} )^\dagger A_{ns}      Y_n^{\rm us}, \qquad
 A_{\nbar s} \to ( Y_n^{\rm us} )^\dagger A_{\nbar s} Y_n^{\rm us}.
\end{equation}
%-----------------------------------------------------------------------------------
 where $Y_{n}^{\rm us}$ is the same as $Y_n$ but with only the ultra-soft gauge fields involved.
After this redefinition, there are no-longer interactions between soft-collinear
and ultra-soft modes. Thus, the original Wilson line operator now separates into
%-----------------------------------------------------------------------------------
\begin{equation}
 Y_{\bar{n}}^\dagger Y_n  \to
 (Y_{\bar{n}}^{\rm sc})^\dagger
 (Y_{\bar{n}}^{\rm us} )^\dagger
 (Y_n^{\rm us} )
 (Y_n^{\rm sc} )
 \label{eq:Yout}
\end{equation}
%-----------------------------------------------------------------------------------
where $Y_{n}^{\text{sc}}$ is a soft-collinear Wilson line, again given
by Eq.~\eqref{Ydef}, but now in terms of decoupled soft-collinear fields. 
%It is this refactorization
%that leads to the refactorization in Eq.~\eqref{factsoft}.
This separation of the soft Wilson line into soft and soft-collinear Wilson lines suggests the refactorized
form in Eq.~\eqref{factsoft}.

Note that this derivation is not a complete proof of the refactorization. We have not given operator definitions for $S_R^\in$ and $S_R^\out$, although
clearly they should be related to matrix elements of the soft-collinear and ultrasoft Wilson lines respectively. This section should be viewed as
an argument for why the soft function might simplify as $R\to 0$. Such simplifications would show up in the calculations of the exact higher-order soft function,
which is not yet known. If the reader does not find this argument convincing, he should ignore this section and consider the phenomenological evidence
for refactorization at small $R$ which we give below.

\subsection{Predictions from refactorization \label{sub:pred}}
To solve the RGE for the soft function, we first extract the anomalous dimensions of the two parts from their NLO expressions.
The anomalous dimension of $S_R^\out(\omega,\mu)$ has no $\mu$ dependence,
consistent with there being no  soft-collinear radiation in the middle region. The $\mu$ dependence of the anomalous dimension
of $S_R^\in(k,\mu)$
is the same as for the thrust soft function.
We can also read off the regular anomalous dimensions at NLO
\begin{align}
\gamma^\out_{S} &= -\gcusp^{(0)} \ln\Rf \\
\gamma^\in_{S} &= \gamma_S  +\gcusp^{(0)} \ln\Rf  \nonumber
\end{align}
where $\gcusp^{(0)} = C_F \alpha_s/\pi$ is the one-loop cusp anomalous dimension.
Generalizing this to higher orders leads to our conjecture Eq.\eqref{eq:outin} with $\gamma_\Delta=0$.
As discussed earlier, we are not interested in non-global logs, so we set $S_f(k_L,k_R,\omega)=1$ for simplicity.

In summary, our conjecture is
\begin{multline}
  S_{R}(k_L,k_R,\omega,\mu) =
  S_{R}^{\in}(k_L,\mu)
  S_{R}^{\in}(k_R,\mu)
  S_{R}^\out(\omega,\mu)
\end{multline}
with $S_R^\in(k,\mu)$ having the $\ln(\mu/k)$ part of the anomalous dimension, and the regular anomalous dimensions
splitting as $\gamma^\out_{S} = -\gcusp \ln\Rf$ and $\gamma^\in_{S} = \gamma_S  +\gcusp \ln\Rf$.

Although we have not proven the soft anomalous dimension splits up in this way at higher order,
we do know that the $R$ dependence must exactly cancel to all orders, which implies it should be universal, naturally suggesting
the cusp anomalous dimension.
As a check, we also know that as $R\to \frac{1}{2}$, $\gamma_\in \to \gamma_S$ and $\gamma_\out \to 0$, since this
is the hemisphere case. At order $\alpha_s^2$, this ${\mathbf \gcusp}$ {\bf ansatz} contributes terms like
$\Gamma_1 \ln\frac{R}{1-R}\ln\tau_\omega$ to the singular $\tau_\omega$ distribution, which we confirm exist by numerical comparison
to QCD (see below). We do not have a proof for the $\gcusp$ ansatz or a strong argument about why  the regular anomalous dimension cannot split
up differently at two loops and beyond.

We also note that the connection between the $\ln R$ terms and the cusp anomalous dimension was pointed out in~\cite{Ellis:2009wj,Ellis:2010rwa}.
In fact, part of the reason we expect that our $\gcusp$ ansatz could hold is related to observations about the structure of the divergences in
the relevant Feynman diagrams which generate the 1-loop anomalous dimensions, first calculated in these papers.
However, without the refactorization at small $R$,
there is no predictive power in this observation. Indeed, generically, there can be an additional anomalous dimension $\gamma_\Delta$,
as indicated in Eq.~\eqref{eq:outin}, and $\gamma_\Delta$ can also depend on $R$.
We are arguing that all the singular dependence on $R$ in the anomalous dimension should be given by the cusp anomalous dimension, so that at small $R$, this
$\gamma_\Delta$ is $R$-independent (and vanishes at one-loop). We cannot rule out such a function which vanishes at $R=\frac{1}{2}$ and goes to a constant
starting at two loops. The small $R$ regime in not mentioned at all in~\cite{Ellis:2009wj,Ellis:2010rwa} (although it is discussed in
a paper \cite{Banfi:2010pa} on non-global logarithms).

As a consequence of refactorization, we can choose separate soft scales for the in-jet and out-of-jet regions, $\mus$ and
$\muw$. Evolving $S_R^\out$ to $\muw$ is straightforward, since there are no scales in its anomalous dimension. Thus
\begin{equation}
S_R^\out(\omega,\mu) = \left(\Rf\right)^{- 2 A_\Gamma(\muw,\mu)}
  S_R^\out (\omega,\muw)\label{sout}  \,.
\end{equation}
The function $A_\Gamma(\nu,\mu)$ is defined in~\cite{Becher:2008cf}.
It follows that choosing $\muw=\omega$ allows us to resum logs of $\omega$, since there is no other scale in this
soft function.

The $S_R^\in$ soft function is similar to the thrust soft function up to $R$-dependence. Its evolution
equation is similarly solved in Laplace space~\cite{Becher:2008cf}. The projection relevant for $\tau_\omega$ adds the $k_L$ and
$k_R$ momentum is just the thrust soft function times a new $R$-dependent factor.

While there may be non-global logs contained in the function $S_f(k_L,k_R,\omega)$ in Eq.~\eqref{factsoft}, for simplicity,
we have set $S_f=1$. Non-global logs are not the subject of this paper. To the extent that our refactorization agrees with the NLO,
it is partly because non-global logs are a numerically small effect in the regime we consider. We refer the reader to ~\cite{Hornig:2011tg},
which appeared after the first version of our paper appeared, and addresses non-global logs for jet thrust and related observables.

Combining the full soft function with the hard and inclusive jet functions gives
the resummed $\tau_\omega$ distribution in SCET
\begin{widetext}
\vspace{-0.3cm}
\begin{multline} \label{scetdist}
\frac{1}{\sigma_0}\left[ \frac{\rd\sigma}{\rd\tau_{\omega}}\right]_{\text{SCET}}=
\exp\Big[ 4S(\muh,\muj)+4S(\mus,\muj)-2A_H(\muh,\mus)+4A_J(\muj,\mus)
\Big]
\left(\Rf\right)^{- 2 A_\Gamma(\muw,\mus)}
\left(\frac{Q^2}{\muh^2}\right)^{-2 A_\Gamma(\muh,\muj)}\\
%-2\ln\frac{Q^2}{\muh^2} A_\Gamma(\muh,\muj) - 2 \ln\Rf A_\Gamma(\muw,\mus\right)] \\
\times
H(Q^2,\muh) S^\out_R(\omega,\muw)\,
\left[\widetilde j\Big( \ln\frac{\mus Q}{\muj^2}+\partial_\eta,\muj\Big)\right]^2\,
{\widetilde s}^\in_{\tau_\omega} \Big(\partial_\eta,\mu_s\Big) \frac{1}{\tau_\omega}
\left(\frac{\tau_\omega Q}{\mus} \right)^{\eta} \frac{e^{-\gamma_E \eta}}{\Gamma(\eta)}\,.
\end{multline}
\end{widetext}

\begin{figure}[t]
\includegraphics[width=0.95\hsize]{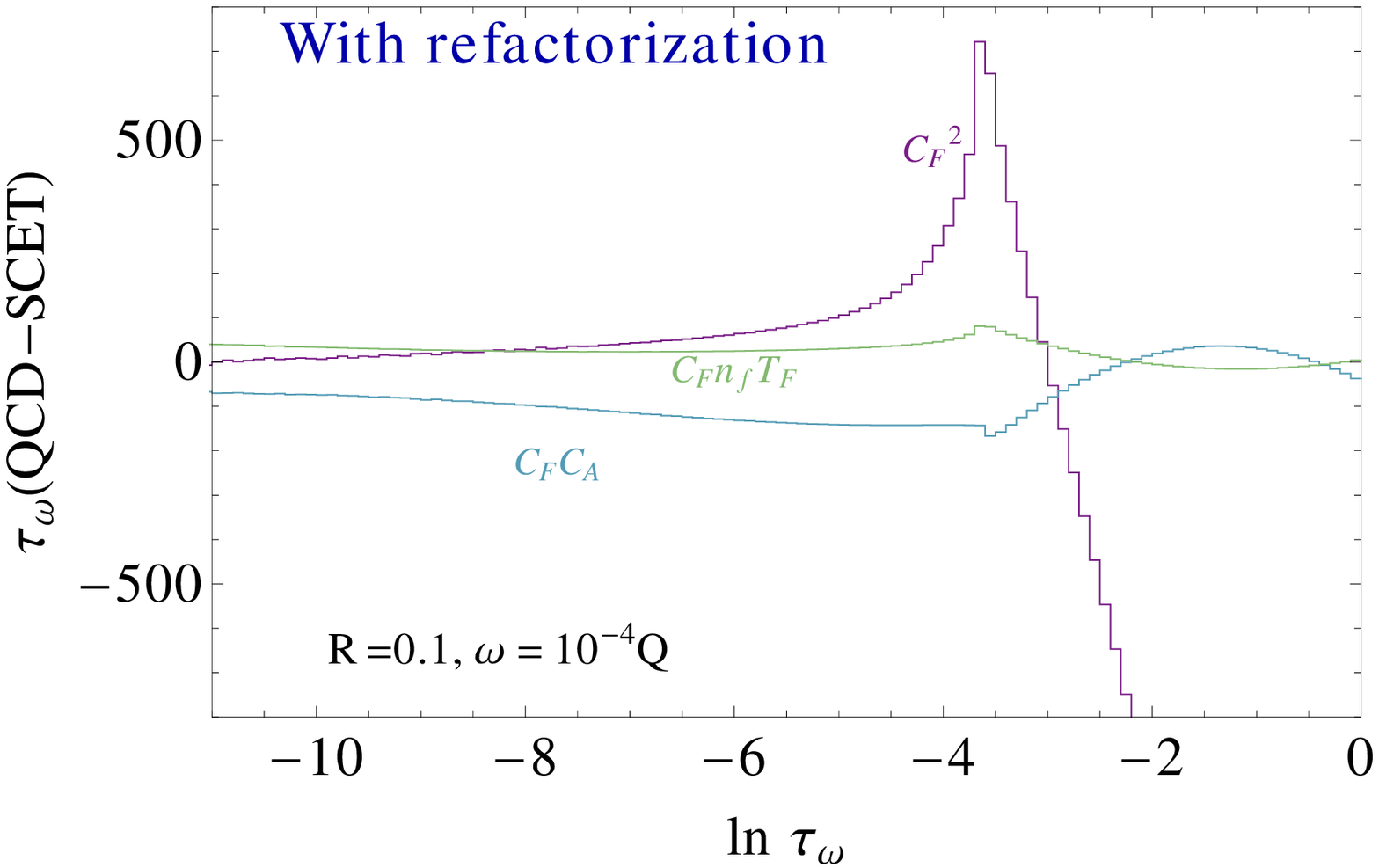}
\caption{The difference between coefficient of $\alpha_s^2$ in $\rd \sigma/\rd \ln \tau_\omega$ in full QCD
and in SCET for $R=0.1$ and $\omega=0.0001 Q$ after refactorization, but not including the $\Gamma_1$ piece.}
\label{fig:diffplotRefact}
% \vspace{-0.4cm}
\end{figure}

\begin{figure}[t!]
\includegraphics[width=0.97\hsize]{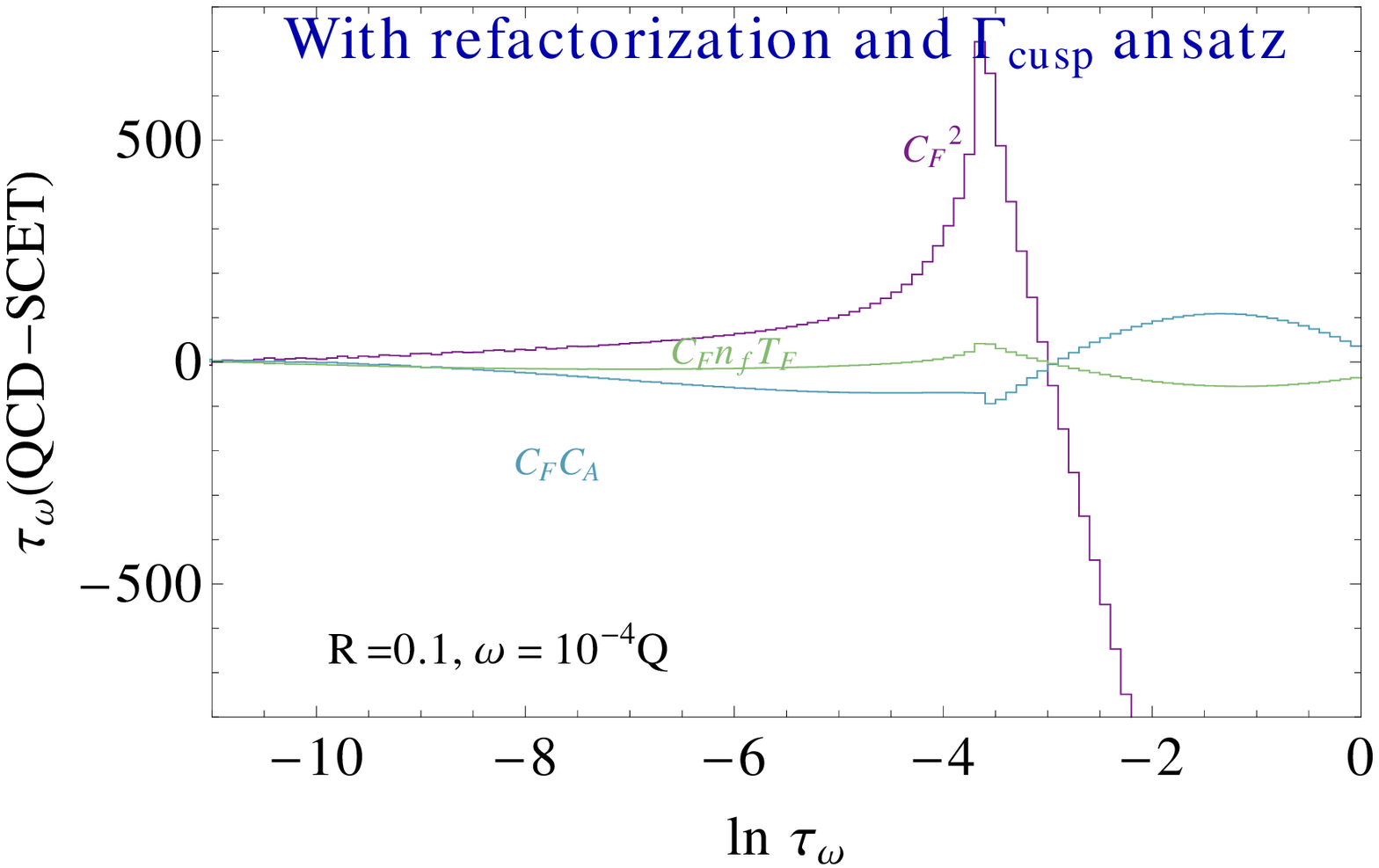}
\caption{The difference between coefficient of $\alpha_s^2$ in $\rd \sigma/\rd \ln \tau_\omega$ in full QCD
and in SCET for $R=0.1$ and $\omega=0.0001 Q$ after refactorization and including the $\Gamma_1$ piece.}
\label{fig:diffplotGamma}
% \vspace{-0.4cm}
\end{figure}

Note that this expression is explicitly independent of $\mu$.
 Demanding independence of the various matching scales fixes the unknown parts of the fixed order soft functions.
Both are of the general form of fixed order expansions in RG-improved perturbation theory, given by $h(L,\mu)$ in
 Eq. (55) of \cite{Becher:2008cf}. In this case
$S^\out_R(\omega,\muw)=h(2\ln\frac{2\omega}{\mu_s},\mu_s)$ with all the $\gcusp$ terms set to zero,
since there are no Sudakov logs outside of the jets, and substituting $\gamma_H \to -\gamma_S^\out$
with $\gamma_S^\out =- \gcusp \ln\Rf$ and
 $c_1^H \to c_1^\out = 2C_F \ln^2\Rf +f_0(R)$.
The in-cone soft function has double logs, and is given by
 ${\widetilde s}^\in_{\tau_\omega}(L,\mu)= h(2L,\mu)$ with $\gamma_H \to -\gamma^\in_{S}$,
$\gamma^\in_{S} = \gamma_H - 2\gamma_J + \gcusp \ln\Rf$, and
$c_1^H \to c_1^\in= -C_F (\pi^2 +2 \ln^2\Rf)$.
Since all the anomalous dimensions are known up to 3-loops, the
first unknown terms are the 2-loop constants in the soft functions.
Thus, the refactorization formula and the $\gcusp$ ansatz, that the 2-loop out-of-cone soft function anomalous dimension is proportional to $\Gamma_1$,
gives us enough for NNLL resummation (again, up to NGLs).

\begin{figure}[t]
\includegraphics[width=0.95\hsize]{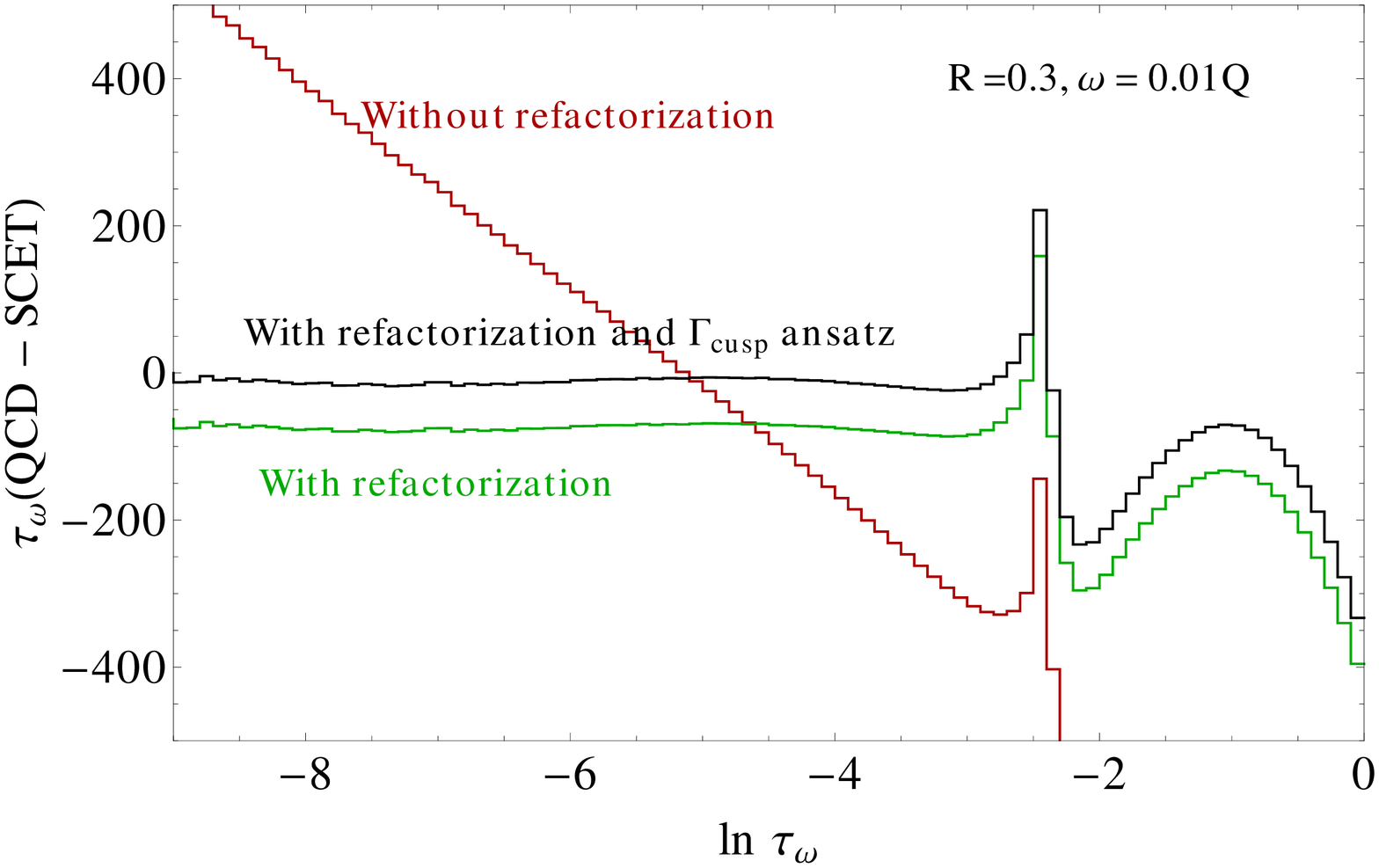}
\caption{The difference between coefficient of $\alpha_s^2$ in $\rd \sigma/\rd \ln \tau_\omega$ in full QCD
and in SCET for $R=0.3$ and $\omega=0.01 Q$ after refactorization and including the $\Gamma_1$ piece.}
\label{fig:diffplotGammaR8}
\vspace{-0.4cm}
\end{figure}

To check our result, we can expand our result to order $\alpha_s^2$ and compare the resulting distributions to full QCD using {\sc event 2}. The results should
agree up to power corrections in $\tau_\omega$, $\omega/Q$ and $R$ and up to NGLs.
Figure~\ref{fig:diffplotRefact} shows the difference between SCET and QCD as a function of
$\ln\tau_\omega$ for $R=0.1$ and $\omega/Q=0.0001$ using refactorization (but setting the 2-loop $S^\out$ anomalous dimension to 0). We see that now both the $C_A C_F$
and $C_F n_f T_F$ color structures go to zero slope, similar to what was seen already in Figure~\ref{fig:diffplotNAE} for the $C_F^2$ color
structure (recall that the $C_F^2$ part of the 2-loop soft function is known exactly from non-Abelian exponentiation). This indicates that we are getting the $\ln\tom/\tom$ terms mostly correct.
We then show in Figure~\ref{fig:diffplotGamma} the further improvement gained by including the contribution from $\Gamma_1$ in the 2-loop soft function anomalous dimension. That the curves now go to zero is a highly non-trivial check. This indicates that we are getting also the $1/\tom$ terms mostly correct. The refactorization
is easiest to confirm in the region with $\tom\sim\omega$, where the NGLs are necessarily small. 

%
%
% \begin{figure}[t]
% \includegraphics[width=0.95\hsize]{ConvPlot.eps}
% \caption{The difference between coefficient of $\alpha_s^2$ in $\rd \sigma/\rd \ln \tau_\omega$ in full QCD
% and in SCET for $R=0.3$ and $\omega=0.01 Q$ using various levels of insight into the SCET soft function.
% If all the terms singular in  $\tau_\omega$ are reproduced, the difference should vanish as $\ln \tau_\omega \to -\infty$.
%  The diagonal red curve shows the prediction if a soft function when only one scale is used (cf. Eq.\eqref{tauwfact}).
% The green curve is the prediction using the factorization of the soft function in Eq.~\eqref{factsoft}.
% The final, black curve also includes the NLO cusp-anomalous dimension terms
% following from the ansatz that $\gamma^\out_{S} =  - \gcusp \ln\frac{R}{1-R}$ to all orders.}
% \label{fig:ConvPlot}
% \vspace{-0.4cm}
% \end{figure}

As discussed above, the refactorization we presented is valid only for small $\omega \lesssim \tom Q$
and small $\tom \ll R \ll 1$, but we expect it to be relevant even for moderate $R$.
In Figure~\ref{fig:diffplotGammaR8} we show the case with $R=0.3$. One can see clearly that the refactorization is
phenomenologically important even for $R$ which is not terribly small.
%
%
% Finally, let us comment on the possibility of non-global logarithms (NGLs). Although the refactorization seems to forbid NGLs, it is hard
% to check numerically whether they are absent.
% Our results hold for $\omega/Q \lesssim \tom \ll R \ll 1$, so one cannot explore the asymptotic region where the NGLs will be large.
% These NGLs would be of the form $\ln\frac{\tom Q}{\omega}$. However, for $\omega/Q \ll \tom$, one cannot explore numerically the region with
% $\tom \to 0$ at fixed $\omega$. For $\tom \ll \omega/Q$, the refactorization is not expected to hold. In the region $\tom \sim \omega/Q$, the NGLs
% are necessarily small.
%
% By direct calculation, using techniques similar to~\cite{Dasgupta:2001sh,Dasgupta:2002bw,Banfi:2010pa}, we calculate the leading non-global log for the $\tom$ soft function in Eq.~\eqref{appngl}. This goes to a constant at $R\to 0$.
% To expain the apprarent inconsistency of this non-global log with our refactorization, we note that the refactorization may be getting right
% only the terms which are singular in $R$, but perhaps not the terms which
% are independent of $R$. This would be consistent with the usual behavior of effective field
% theories which reproduce only singular behavior. Indeed, the constant terms in $R$ would become $\delta(R)$ in a differential distribution. However,
% for $R=0$, $\omega$ is necessarily larger than $\tau$, so the derivation of the refactorization breaks down.
%  Understanding these issues in more detail is an important area for future considerations.

\section{Conclusions}
In this paper we have considered a number of important issues which will be relevant to producing accurate distributions of jet
masses and other jet substructure observables at colliders. We first considered the possibility of an inclusive jet mass measurement,
such as the asymmetric thrust observable. In this case, we were able to resum logarithms of the jet mass to next-to-next-to-leading
logarithmic accuracy. These distributions, when expanded to order $\alpha_s^2$ showed excellent agreement with numerical calculations in full
QCD in the threshold region.

At hadron colliders, it will be difficult to use an inclusive jet mass observable. Instead, one must restrict consideration to the hardest jet
or the hardest pair of jets in the event. In this case, it is important to veto additional jets, which introduces a veto scale $\omega$. Then
there are at least three parameters, $\omega/Q$, where $Q$ is some hard energy scale, the jet shape $\tau$ and the jet size $R$. The goal
in resumming the jet shape would be to predict coefficients of $\alpha_s^i\ln^j\tau$ without doing the full calculation to $i$th order in perturbation theory.
In general, this is difficult, but we argued that there is a refactorization which simplifies the calculation for the observable jet thrust, $\tom$.
Jet thrust is defined as the sum of the jet masses squared normalized to the center of mass energy $Q$, after a vetoing the energy $\omega$ of the third hardest jet.
We argue that the refactorization holds in the limit $\omega/Q \lesssim \tom \ll R \ll 1$.
We presented kinematic scaling arguments, using the language of Soft-Collinear Effective Theory,
to justify the refactorization.

In order to test the refactorization, we compared the predictions to the exact NLO event shape distributions in full QCD. This provided very strong numerical evidence
that the refactorization correctly predicts the dominant part of the jet mass distributions. We further argued that the $R$-dependence may be universal, and therefore
may depend on the cusp anomalous dimension. This $\gcusp$ ansatz was also confirmed by comparison to full QCD.
The predictions from refactorization are independent of the separate, and important, issues of non-global logs.
NGLs are apparently a numerically small correction to our results, but are not the focus of this paper.
Extrapolating away from $R=0$ our result still provides
good agreement with full QCD.  Thus the small $R$ expansion seems like a productive direction for further investigation.

The improved agreement with full QCD after refactorization is impressive, but still not completely well understood. It would be helpful to have the full soft function calculated, along the lines of~\cite{Kelley:2011ng,Hornig:2011iu}, to understand exactly which terms are reproduced and which are not. It would also be helpful to have
gauge-invariant operator definitions of the functions into which the soft function factorizes, so that they can be computed directly. Progress along these lines has
already appeared, in~\cite{ninja}. Most importantly, it will be critical to construct jet mass observables for hadron colliders which can be computed and resummed to
next-to-next-to-leading logarithmic accuracy so that direct comparisons to collider data can be made.

\vskip0.5cm
\section*{Acknowledgments}
HXZ would like to express his gratitude to Chong Sheng Li
for his support and useful discussion on the project.
The authors would also like to thank I.~Stewart and C.~Lee for helpful conversations.
This research was supported in part by the Department of Energy,
under grant DE-SC003916 and the National Natural Science Foundation of China, under Grants No.11021092 and No.10975004.
\vskip1cm

%%%%%%%%%%%%%%%%%%%%%%%%%%%%%%%%%%%%%%%%%%%%%%%%%%%%%%%%%%%%%%%%%%%%%%%%%%%%%%%%%%%%%%%%%%%%%%%%%%%%%%%%%%%%%%%%%%%%%
%%%%%%%%%%%%%%%%%%%%%%%%%%%%%%%%%%%%%%%%%%%%%%%%%%%%%%%%%%%%%%%%%%%%%%%%%%%%%%%%%%%%%%%%%%%%%%%%%%%%%%%%%%%%%%%%%%%%%
%%%%%%%%%%%%%%%%%%%%%%%%%%%%%%%%%%%%%%%%%%%%%%%%%%%%%%%%%%%%%%%%%%%%%%%%%%%%%%%%%%%%%%%%%%%%%%%%%%%%%%%%%%%%%%%%%%%%%
\appendix
\begin{widetext}
\section{NNLO expansion of SCET results}
\label{app:expansion}
In this appendix, we give the $\tom$ distribution in SCET to order $\alpha_s^2$.  The SCET distribution at order $\alpha_s$ is given in Eq.~\eqref{tomSCET}.
At order $\alpha_s^2$, we can write
\begin{multline}
\frac{\tom}{\sigma_0}\left[\frac{\rd \sigma}{\rd \tom}\right]_{\text{SCET}}=
\left(\frac{\alpha_s}{4\pi}\right)^2 \Big\{ C_F^2\Big[ f_0^{C_F}+ f^{C_F}_{\text{NAE}} \Big]
%\\
+C_F C_A \Big[ f_0^{C_A}  + f_{\Gamma_1}^{C_A} \Big]
+C_F n_f T_F \Big[ f_0^{n_f}  + f_{\Gamma_1}^{n_f} \Big]
%\\
+ C_F \beta_0 f_{\text{Refact}} \Big\}
\end{multline}
where $\beta_0 = \frac{11}{3}C_A -\frac{4}{3} n_f T_F$ is the leading $\beta$-function coefficient in QCD.

The parts coming from the 1-loop result and the original SCET factorization formula for $\tom$ are
\begin{multline}
 f_0^{C_F}
=32 \ln^3\tom+\ln^2\tom \left[72-96 \ln \Rf\right]\\
 +\ln\tom \Big[32 \text{Li}_2(1-R)-32 \text{Li}_2 R
+\ln \Rf \left(64 \ln \frac{2 \omega }{Q} + 32\ln R -96 \right)-16 \pi ^2+52\Big]\\
+24 \text{Li}_2(1-R)-24 \text{Li}_2(R)
+\ln \Rf\left(48 \ln\frac{2 \omega }{Q}+ 24\ln R + 24\pi^2 - 16\right)+16 \zeta_3-8 \pi ^2+9 
\end{multline}
\begin{multline}
f_0^{C_A}
= 44 \ln^2\tom + \ln\tom \left[-\frac{352}{3} \ln\Rf+\frac{8}{3} \pi^2-\frac{338}{9}\right]\\
+\frac{176}{3}\ln\frac{2\omega}{Q} \ln\Rf+\frac{88}{3} \text{Li}_2(1-R)-\frac{88}{3} \text{Li}_2 R+\frac{88}{3} \ln\Rf\ln R
+24 \zeta_3 -57 
\end{multline}
\begin{multline}
f_0^{n_f}
= -16 \ln^2\tom+\ln\tom \left[\frac{128}{3} \ln\Rf+\frac{88}{9}\right] \\
-\frac{64}{3} \ln\frac{2\omega}{Q} \ln\Rf - \frac{32}{3} \text{Li}_2(1-R) + \frac{32}{3} \text{Li}_2 R -\frac{32}{3} \ln\Rf \ln R+20 
\end{multline}
The part coming from non-Abelian exponentiation is
\begin{multline}
 f_{\text{NAE}}^{C_F}=
64 \ln^2\Rf\ln\tom +32 \ln\Rf \left(2 \text{Li}_2 R -2 \ln \Rf \ln\frac{2\omega}{Q} - \ln\Rf\ln (4 R) + \ln (1-R)\ln R - \frac{5\pi^2}{12}\right) 
%-32 \ln\Rf \ln R\\
%+32 \ \ln (1-R) \ln R-64 \ln\Rf \ln\frac{\omega}{2Q}-\frac{40 \pi^2}{3} \Big]
\end{multline}
The part from the soft-refactorization is
\begin{equation}
f_{\text{Refact}}
= 16 \ln\frac{\tom Q}{2 \omega} \ln \Rf + 4 \text{Li}_2(R)
+\ln^2(1-R) + 2\ln (1-R) \ln R - \ln^2 R
-\frac{\pi ^2}{3}. 
\end{equation}
The parts from the $\Gamma_1$ ansatz are
\begin{equation}
f_{\Gamma_1}^{C_A} =\frac{8}{9} \left(67-3 \pi ^2\right) \ln\Rf, \qquad
f_{\Gamma_1}^{n_f} =-\frac{160 }{9}\ln\Rf .
\end{equation}
Adding these pieces, the complete $C_F C_A$ color structure at order $\alpha_s^2$ is relatively simple
\begin{multline}
 f^{C_A} = 44 \ln^2\tom+\left(\frac{8 \pi^2}{3}-\frac{176}{3}\ln\Rf-\frac{338}{9}\right)\ln\tom
%\\
+\frac{536}{9}\ln\Rf+ \frac{44}{3} \ln^2\Rf-\frac{8 \pi^2}{3}\ln\Rf+24 \zeta_3-57 
\end{multline}
Note the $\omega$ dependence has completely dropped out of the sum for the $C_F C_A$ (and the $C_F n_f T_F$) color structure
order $\alpha_s^2$. This follows simply from the refactorization. There is $\omega$ dependence in the $C_F^2$ terms.

By explicit calculation, we can also work out the non-global log in the $C_F C_A$ color structure.
Rather than use the Cambridge/Aachen algorithm, for simplicity, we calculate the contribution by just including all radiation within $R$
of the jet direction. This produces a non-global log of the form
\begin{equation}
f^{C_A}_{\rm NGL}=\left(- \frac{16\pi^2}{3}+32\mathrm{Li}_2
\frac{R^2}{(1-R)^2}\right)\ln\left(\frac{\tau_\omega
Q}{2 \omega}\right). \label{appngl}
\end{equation}
This vanishes at $R=\frac{1}{2}$ and goes to $-\frac{16\pi^2}{3}\ln\frac{\tom Q}{2\omega}$ at $R=0$.

\vskip2cm
\end{widetext}

%\pagebreak

\end{document}